\documentclass[10pt]{article}
\usepackage{graphicx}
\usepackage{epsfig}

\def\b{{{ b}}}
\def\s{{{ s}}}
\def\c{{{ c}}}
\def\d{{{ d}}}
\def\u{{{ u}}}
\def\q{{{ q}}}
\def\Q{{{ Q}}}
\def\g{{{ g}}}
\def\p{{{ p}}}
\def\n{{{ n}}}
\def\e{{{ e}}}
\def\D{{{ D}}}
\def\B{{{ B}}}
\def\T{{{ T}}}

\title{The $\T_{\c\c}=\D\D^*$ molecular state}
\author{D. Janc$^1$ and M. Rosina$^{1,2}$\\
{\sl $^1$ Jo\v zef~Stefan Institute, Jamova 39, P.O.~Box 3000, SI-1001 Ljubljana, Slovenia}\\
{\sl $^2$ Faculty of Mathematics and Physics, University of Ljubljana}}
\begin{document}
\date{}
\maketitle

\begin{abstract}
We show that the molecule-like configuration of $\D\D^*$
enables weak binding
with two realistic potential models
(Bhaduri and Grenoble AL1).
Three-body forces may increase the binding and
strengthen the $\c\c$ diquark configuration. As a signature
we propose the branching ratio between radiative and pionic decay.
\end{abstract}

\section{Introduction }

The motivation to study tetraquarks (also called dimesons)
comes from our curiosity whether or not we can extrapolate our
understanding of mesons and baryons (in terms of quark models)
to two-hadron systems. Dimesons are simpler than dibaryons
(or nuclear forces) since they represent a four-body rather
than a six-body system. Heavy dimesons are cleaner than light ones
since nonrelativistic parameterization and treatment are
more justified and since they are likely to be longlived.
Therefore in this paper we study double heavy tetraquarks
as prototypes. 
In our extrapolation from mesons and baryons to tetraquarks
we assume the "$\lambda\cdot\lambda$" type of colour dependence 
of the interaction. It seems to work properly when going from mesons to
baryons. It is a challenge to use tetraquarks as a test whether or not
this assumption is valid also for larger systems.

However, the question arises whether or not there is a signature of tetraquarks
(dimesons) which could clearly distinguish their decay from the decay
of two independent mesons.
Consider the analogy. The mass of the neutron is 1.3 MeV
larger than the mass of the proton,
making the neutron unstable against the weak interaction and results in
the $\n\to\p \e^-\bar{\nu_\e}$ decay. But when the neutron is bound
in the deuteron with a binding energy of -2.23 MeV,
the decay is kinematically forbidden and the neutron becomes
stable. If we replace the two baryons in the deuteron with two mesons
we obtain the dimeson. When one of the mesons is a vector
meson, its dominant decay mode in the case of weak binding would be
the radiative decay $\bar{\B}^*\to \bar{\B}\gamma$, or in
the system of the $\D$ mesons $\D^*\to \D\gamma$,
as well as the strong decay $\D^*\to \D\pi$.
The $\bar{\B}^*\bar{\B}$ dimeson, however, is probably bound strongly
enough so that radiative decay becomes energetically forbidden
\cite{JR,SB,BS} and it can decay only weakly. The binding of the
$\D^*\D$ dimeson is expected to be much weaker, but it might still
be stable against strong interaction so as to decay only electromagnetically,
or, at least, the strong decay might be considerably suppressed.
The dramatic change
of the decay modes and lifetime of vector mesons can serve as
a tool for detecting tetraquarks with nonzero total spin. This effect
would be very helpful in situations where the binding energy is small
compared to experimental errors so that the detection of the tetraquark
from the invariant mass of final particles could not distinguish between
events where the two initial mesons were either free
or weakly bound before the decay.

In this study we consider tetraquarks made up of the same type of mesons, 
namely $\D\D^*$ and
$\bar{\B}\bar{\B}^*$ tetraquarks with quantum numbers $I=0,S=1,P=+1$,
since they are the best candidates for binding
with respect to the $\D\D^*$ ($\bar{\B}\bar{\B}^*$) threshold,
as was already noted previously \cite{JR,ZSB,SB,SBii,BS}.
Such states cannot decay strongly or electromagnetically
into two $\bar{\B}$ or two $\D$ mesons in the S wave due to angular momentum
conservation nor in the P wave due to parity conservation.

There are two extreme spatial configurations of quarks in a tetraquark.
The first configuration which we call {\em atomic}
is similar to $\bar{\Lambda}_\b$, with a compact
$\b\b$ diquark instead of $\bar{\b}$,
around which the light antiquarks are in a similar state as in the
$\bar\Lambda_\b$ baryon. 
The stability of the double heavy tetraquarks 
with this structure was first investigated by
Lipkin \cite{lipkin}.  
The second configuration which we call {\em molecular}
is deuteron-like, where two heavy
quarks are well separated and the light antiquarks are bound to them
similarly as in the case of free mesons. This configuration is more likely
to appear in weakly bound systems and was studied by Manohar and Wise \cite{wisse}
and T\"ornqvist \cite{torn1, torn2}
in the framework of the pion exchange between two heavy mesons.
It has been shown that nonrelativistic
potential models in general give rise to {\em atomic} structure for the
$\bar{\B}\bar{\B}^*$ tetraquark \cite{JR,SB}, due to the large quark mass
asymmetry.
We use the $\bar{\B}\bar{\B}^*$ system as a benchmark against which
we compare other tetraquarks.

In Sect. 2 we repeat calculations of the $\bar{\B}\bar{\B}^*$ tetraquark
with the Bhaduri potential \cite{Bh}
and also with the AL1 potential \cite{al1}
which due to the additional mass-dependent smearing of
the spin-spin interaction gives a better description of meson spectroscopy.
We then present results for the $\D\D^*$ tetraquark, which is on the
verge of being either bound
or a resonant state, depending on the effective potential used in
the calculations. We show that its structure is {\em molecular}, but
with the introduction of the three-body force (Sect. 3) it can become {\em atomic}.
The fact that this system is so close to the $\D+\D^*$ threshold makes it
very sensitive to the details of the effective interaction and therefore
a promising candidate for studying the nature of the effective interaction
between constituent quarks in nonrelativistic potential models.
The estimated production rate is not high but tolerable.
Due to the strong influence of weak binding on the decay channels, it
presents a very interesting experimental situation (Sect. 4).

\section{Bound states of heavy tetraquarks}

%The general idea of possible stable dimesons (tetraquarks) with two
%heavy quarks and two light antiquarks has been first suggested by [Jaffe??].
%It has also been proposed by [Ader] and [Heller] where the binding energy
%as a function of the mass ratio has been estimated. The $b/u$ mass ratio
%is shown to be much larger than critical making $T_{bb}$ stable,
%while the $c/u$ mass ratio is just about critical.
%
The general idea of possible stable heavy tetraquarks 
has been first suggested by Jaffe \cite{jaffe}.
It soon became clear that the systems with unequal
masses of the quarks and the antiquarks are more promising,
since the binding energy strongly depend on the mass ratio 
\cite{ader,heller1,heller2}. The $b/u$ mass ratio
is shown to be large enough to make $T_{bb}$ stable,
while the $c/u$ mass ratio is under-critical for atomic structure.
The  calculation 
of various tetraquarks
in the harmonic oscillator basis \cite{SB, SBii} have shown
that only two  tetraquark systems
have their energy lower than the two-free-meson threshold, namely
$\b\b\bar{\u}\bar{\d}$ (I=0,J=1) which we denote as $\T_{\b\b}$ and
$\b\b\bar{\s}\bar{\u}$ or $\b\b\bar{\s}\bar{\d}$ (I=1/2,J=1), while
$\b\c\bar{\s}\bar{\u}$ or $\b\b\bar{\s}\bar{\d}$ (I=1/2,J=1)
lie on the threshold.
For deeply bound states, these results should be very accurate
but since this basis cannot accommodate asymptotic states of two
free mesons, there is an open question whether or not
weakly bound states of two mesons have been missed.

In our work we use an expansion in the basis proposed by \cite{BS} which
is described in Appendix A. There are three different sets of internal
coordinates. The first one (Fig.\ref{ks} a)) is convenient for expansion of
those strongly correlated and deeply bound tetraquarks where we expect
the {\em atomic} structure in which the diquark
in $\T_{\Q\Q}$ formed by two
heavy  quarks plays a similar role as the heavy $b$ antiquark in
the $\bar{\Lambda}_\b$ baryon, while the light quarks in both systems 
are in the same radial, spin, colour and isospin configurations.
The second and third sets from Fig. \ref{ks} represent
the direct and exchange meson-meson channels. These configurations are needed
to build up the basis for the two free mesons - the threshold state,
and are also of crucial importance for searching for weakly bound
tetraquarks where the molecular structure would be dominant.

We search for eigenstates of our Hamiltonian using the variational method,
applying a general diagonalization
of the Hamiltonian (Appendix B) spanned by the non-orthogonal basis functions
constructed in Appendix A.
We built the basis functions step by step by adding the best
configurations from Fig. \ref{ks} with the best colour-spin configurations
allowed for our quantum numbers (IS=01, positive parity and colour singlet),
after having optimized the corresponding Gaussian widths.
In order to obtain a 0.1 MeV accuracy we constructed bases in this way
with up to $N_{max}=90$ and  $N_{max}=140$ functions for the
$\T_{\b\b}$ and  $\T_{\c\c}$ tetraquarks, respectively (Appendix D).
These basis states can also accommodate two asymptotically
free mesons if the four-body problem has no bound state.

\subsection{$\T_{\b\b}$}

First we test our method on the $\T_{\b\b}$ system and compare our results
with Ref. \cite{SB, SBii}. In our calculations we use two one-gluon exchange
potentials, the Bhaduri and AL1 potential.
Their properties are described in Appendix B.
The Bhaduri potential quite successfully describes the
spectroscopy of the meson, as well as baryon ground states. This is an
important condition since in the tetraquarks we have both quark-quark and
quark-antiquark interactions. The AL1 potential is just an improvement
of the Bhaduri potential where the smearing of the colour-magnetic term
in the Hamiltonian depends on the masses of the quarks. This then results 
in better quality of the meson spectra, in particular in the charmonium
sector, where the Bhaduri potential predict only half of the observed
hyperfine splitting between the $\eta_{c}$ and $J/\psi$ state.

The results obtained with the Bhaduri potential are presented
in the first three columns of
 Table \ref{tabelaBB} where they are compared with results from
\cite{SB}. In the last three columns our and \cite{SBii}
results for the AL1 potential are listed.
Since the harmonic oscillator basis cannot accommodate two asymptotically
free mesons, one obtains a positive binding energy, as for example for
spin 1, isospin 1 state calculated with the Bhaduri potential.

\begin{table}[here]
\caption{The mass of the $\T_{\b\b}$ tetraquark. Column 1: spin S,
isospin I, Column 2: lowest meson-meson threshold for the Bhaduri potential,
Column 3: our results, Column 4: results of ref.\cite{SB} where 
expansion in the harmonic oscillator basis was used, 
Column 5: lowest meson-meson threshold
for the AL1 potential, Column 6: our results with the AL1 potential,
Column 7: results of ref.\cite{SBii} where the same basis was used as in
ref.\cite{SB}.}
\label{tabelaBB}
\vspace {0.5cm}
\begin{tabular}{c c c c c c c}
\hline
IS  & threshold & N$_{max}$=90 & Ref.\cite{SB} &
threshold &N$_{max}$=90 & Ref.\cite{SBii} \\
& [Bh] & [Bh] & [Bh] & [AL1] & [Al1] & [AL1]\\
\hline
01 &   10650.9 &  10518.9 & 10525  & 10644.1  &  10503.9 & 10509\\
10 &   10601.4 &  10601.4 &$>$10642  & 10587.0   & 10587.0 & --\\
11 &   10650.9 &  10650.9 & 10712  & 10644.1  &  10644.1 & --\\
\hline
\end{tabular}
\end{table}

\begin{figure}[here]
\centering
\epsfig{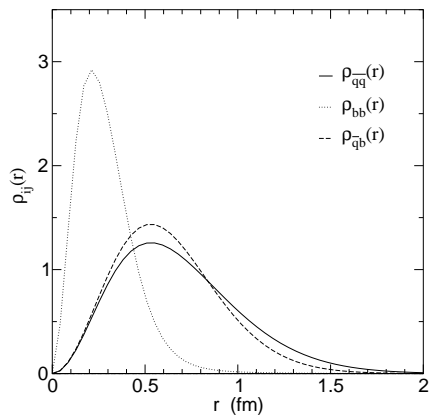}
\caption{%
Probability density of two heavy quarks $\rho_{\b\b}$,
of two light antiquarks $\rho_{\bar q\bar q}$ and of a light antiquark and
heavy quark $\rho_{\bar q b}$ in $\T_{\b\b}$ as a function of interquark
distances for the AL1 potential ($q$ is $u$ or $d$).
}
\label{slbb}
\end{figure}
\newpage

\begin{figure}[h h h]
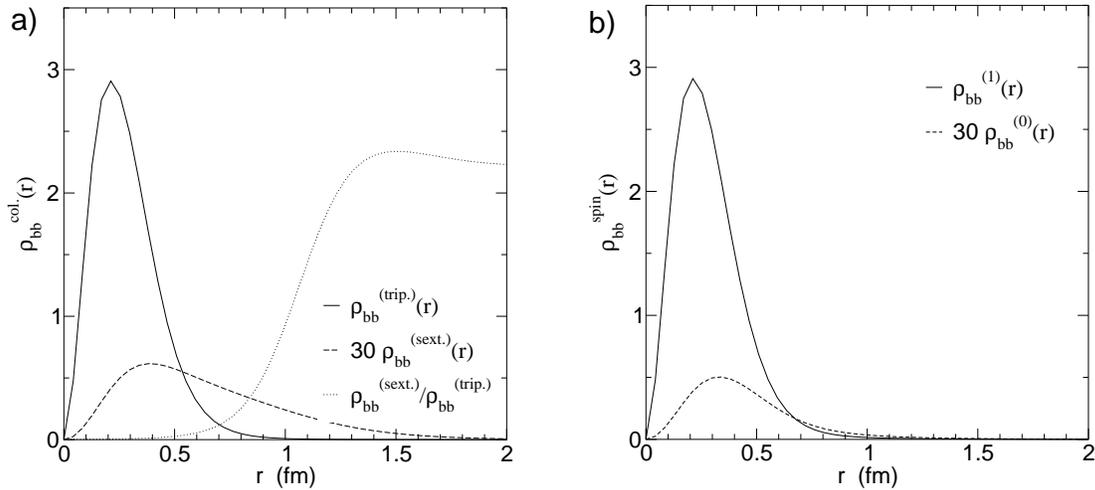

\vspace{3mm}
\includegraphics[width=0.45\linewidth]{vfBBn80color.eps}
\hspace{0.05\linewidth}
\includegraphics[width=0.45\linewidth]{vfBBn80spin.eps}
\caption {%
Results of calculations with the AL1 potential.
{\bf{a)}}: Probability densities  $\rho_{\b\b}^{\mathrm{(trip.)}}$ 
and $\rho_{\b\b}^{\mathrm{(sext.)}}$
of the two heavy quarks
projected on the colour triplet
$|\bar 3_{12}3_{34}\rangle_C$ and colour sextet states
$|\bar 3_{12}3_{34}\rangle_C$, respectively, and their ratio.
{\bf{b)}}: Probability density  $\rho_{\b\b}^{(1)}$ and  $\rho_{\b\b}^{(0)}$
of the two heavy quarks
projected on the spin 1 states
$|1_{12}0_{34}\rangle_S$ and spin 0 states
$|0_{12}1_{34}\rangle_S$, respectively.
}
\label{slbbspincolor}
\end{figure}

To obtain a better understanding of the $\T_{\b\b}$ tetraquark we now turn
to the radial, spin and colour structure of this system.
The probability densities for finding quark $i$ and
(anti)quark $j$ at the interquark distance $r_{ij}$
as shown in Fig. \ref{slbb} is calculated via
$$\rho_{ij}(r)=
\langle \psi|\delta(r-r_{ij})|\psi\rangle.$$
The projection of this probability density on the colour triplet states
$|\bar3_{12}3_{34}\rangle_C$ is 
$$\rho_{ij}^{\mathrm{\mathrm{(trip.)}}}(r)=
\langle \psi|\bar 3_{12}3_{34}\rangle_C \langle \bar
3_{12}3_{34}|_C \delta(r-r_{ij})|\psi\rangle,$$
and similarly for the other spin and colour projections presented
in Fig. \ref{slbbspincolor}.

From Fig. \ref{slbb} it can be seen that the  probability
density of two heavy b quarks is strongly localized
so one can make a rough approximation in which the heavy
diquark is pointlike. The average distances between
a light antiquark and a heavy quark, and between
a light antiquark and a light antiquark are almost the same, so the heavy
diquark and the two light antiquarks form some sort of equilateral
triangle which can also be seen as a "Y" shape configuration.

In Fig. \ref{slbbspincolor}a we see that the dominant colour configuration is
$\bar 3_{12}3_{34}$ where two heavy quarks are in the colour antitriplet state.
The $6_{12}\bar 6_{34}$ configuration becomes relatively important at very
large distances where the absolute probability density is negligible.
The ratio of these two configurations approaches 2 for large separation
between two b quarks which is consistent with the fact that for large distances
we have two white mesons and the octet configuration is not present.
This can be seen from the decomposition of colour octet and singlet 
(quark-antiquark)-(quark-antiquark) states into colour sextet and triplet 
diquark-antidiquark states.
\begin{eqnarray}
|1_{13} 1_{24}\rangle_C =&\sqrt{\frac{1}{3}}|\bar{3}_{12} 3_{34}\rangle_C+
\sqrt{\frac{2}{3}}|6_{12} \bar{6}_{34}\rangle_C,\nonumber\\
|8_{13} 8_{24}\rangle_C =&-\sqrt{\frac{2}{3}}|\bar{3}_{12} 3_{34}\rangle_C+
\sqrt{\frac{1}{3}}|6_{12} \bar{6}_{34}\rangle_C.\nonumber\\
|1_{14} 1_{23}\rangle_C =&-\sqrt{\frac{1}{3}}|\bar{3}_{12} 3_{34}\rangle_C+
\sqrt{\frac{2}{3}}|6_{12} \bar{6}_{34}\rangle_C,\nonumber\\
|8_{14} 8_{23}\rangle_C =&\sqrt{\frac{2}{3}}|\bar{3}_{12} 3_{34}\rangle_C+
\sqrt{\frac{1}{3}}|6_{12} \bar{6}_{34}\rangle_C.\nonumber
\end{eqnarray}
Since the heavy diquark is in a spatial symmetric and colour antitriplet state,
it must be due to the Pauli principle in the spin symmetric S=1 state. This can
also be seen in Fig. \ref{slbbspincolor}b. In this figure the
$|1_{12}1_{34}\rangle_S$
configurations are not shown since they are three orders of magnitude smaller
and  can thus be neglected.

We have shown that the $\T_{\b\b}$ tetraquark has an {\em atomic}
$\bar{\Lambda}_\b$-like structure where the
heavy diquark in the colour antitriplet state can be approximated with a heavy
antiquark $\bar{\b}$, while the light antiquarks are in isospin 0 
and spin 0 state
just like in $\bar{\Lambda}_\b$ baryons. This then justifies
the assumptions made in \cite{JR}
that  the light quarks in $T_{bb}$ have a similar spatial function 
as in $\Lambda_b$. The binding energy can  then be
phenomenologically estimated \cite{JR} to be -134 MeV for
the AL1 potential and -139 MeV for the Bhaduri potential,
which is close to the detailed calculations presented here.
This result is independent of the form of the
interaction between $\bar{u}$ and $\bar{d}$ since if the
$ud$ wave function in $T_{bb}$ and $\Lambda_b$ is the same, then also the
interaction matrix element is the same in both cases and cancels out
in the comparison. For this reason, $T_{bb}$ with its $\Lambda_b$-like
structure cannot distinguish between one-gluon (OGE) \cite{Bh, al1} and Goldstone 
boson exchange (GBE) \cite{graz}. 
As argued in Appendix C, it is crucial to have correct $\Lambda_b$
mass to ensure the correct $ud$ interaction and the GBE calculations which
do not fit the $\Lambda_b$ mass 
\cite{vijande, pepin} may strongly
overbind the $T_{bb}$ as well as $T_{cc}$. 

Because of a dominant colour triplet-triplet and spacial
"Y" shape structure of the tetraquark,
the introduction of a weak colour dependent three-body
interaction would only shift the
mass of the tetraquark and not produce any significant changes in the
wave function, similarly as in the baryon sector.
This effect can be compensated by a reparametrisation
of parameters in the two-body potential or constituent quark masses
so as to reproduce correct baryon ground states; the shrinking of
baryon spectra due to three-body interaction does not concern
our considerations. Therefore the $T_{bb}$ tetraquark is unsuitable
for
studying the influence of the three-body interaction. 

There are also experimental problems with the $\T_{\b\b}$ tetraquark.
The only promising production mechanism -- double two-gluon fusion
$(\g+\g\to \b+\bar{\b})^2$ -- gives a very small production rate of about 5
events per hour at LHC \cite{FT}, \cite{fb2002}. Moreover, since  $\T_{\b\b}$
is below the total mass of two $\bar{\B}$
mesons, it can decay only weakly and thus has no
characteristic decay different from separate $\bar{\B}$ decays.
This then also makes the  $\T_{\b\b}$ tetraquark
unpromising from the experimental point of view.

\subsection{$\T_{\c\c}$}

The $\T_{\c\c}$ tetraquark is much more promising than the
$\T_{\b\b}$ tetraquark. It can be more easily produced and detected
(Sect. 4) and we shall see that it better discriminates between
different binding mechanisms.

With the expansion of the tetraquark wave function in the 
harmonic oscillator basis
one cannot find any bound state for
the $\T_{\c\c}$ system with the Bhaduri or AL1 potential. But as mentioned
in the previous subsection, this can also be due to the fact that this
method can miss weakly bound states. And this is exactly what happened
as one can see from Table \ref{tabelaDD} where our results
are presented. With both potentials, 
a weakly bound state does appear.

\begin{table}[here]
\caption{The mass of the $\T_{\c\c}$ (S=1, I=0) tetraquark. Column 1: type
of potential, Column 2: lowest meson-meson threshold for a given potential,
Column 3: our results, Column 4: results of ref.\cite{SB, SBii} where
expansion in the harmonic oscillator basis was used.}
\label{tabelaDD}
\vspace {0.5cm}
\begin{tabular}{c c c c }
\hline
  & threshold & N$_{max}$=140& Ref.\cite{SB} \\
\hline
Bhaduri &   3905.3 &  3904.7& 3931\\
AL1 &   3878.6 &      3875.9&  3892\\
\hline
\end{tabular}
\end{table}

\begin{figure}[here]
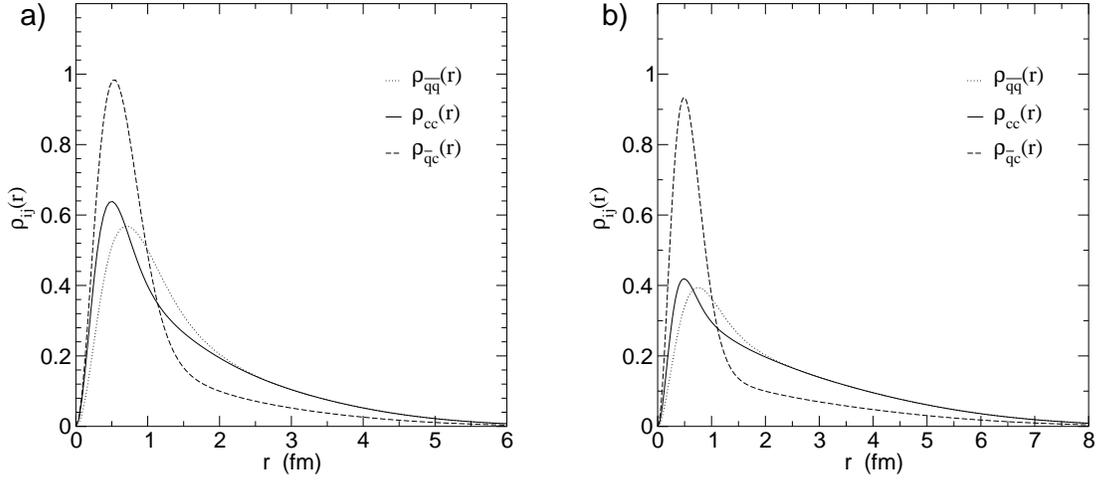

\includegraphics[width=0.45\linewidth]{vfDDn80.eps}
\hspace{0.05\linewidth}
\includegraphics[width=0.45\linewidth]{Bh-qqcc-N140.eps}
\caption {%
Probability density of the two heavy quarks $\rho_{\c\c}$,
of the two light antiquarks $\rho_{\bar q\bar q}$ and of a light antiquark and
a heavy quark $\rho_{\bar q c}$ in $\T_{\c\c}$ as a function
of the interquark distance. {\bf{a)}}: results for the AL1
potential. {\bf{b)}}: results for the Bhaduri potential
}
\label{slcc}
\end{figure}

\begin{figure}[ht!]
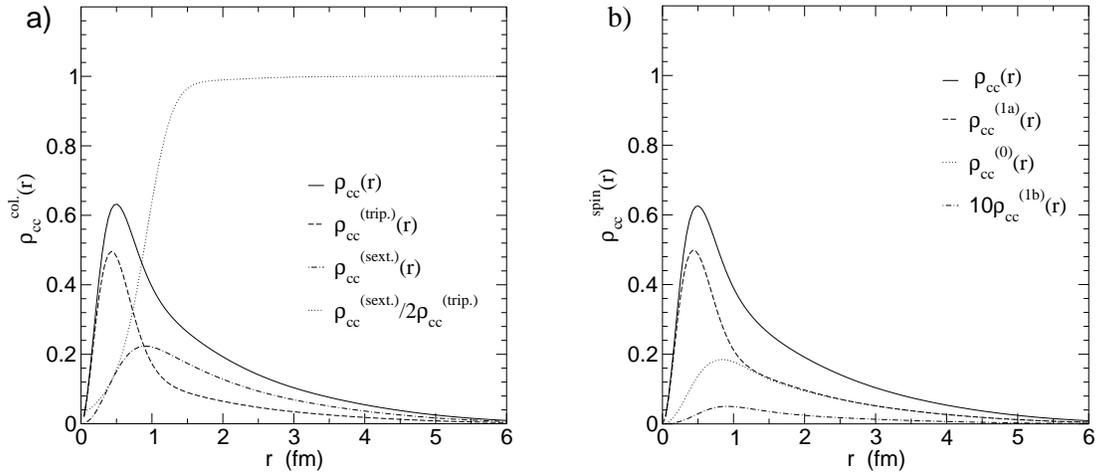

\includegraphics[width=0.45\linewidth]{vfDDn99color.eps}
\hspace{0.05\linewidth}
\includegraphics[width=0.45\linewidth]{vfDDn99spin.eps}
\caption {%
Results of calculations with the AL1 potential.
{\bf{a)}}: Probability densities $\rho_{\c\c}^{\mathrm{(trip.)}}$ and 
$\rho_{\c\c}^{\mathrm{(sext.)}}$
of the two heavy quarks as a function of
interquark distance
projected on the colour triplet state
$|\bar 3_{12}3_{34}\rangle_C$ and on the colour sextet state
$|\bar 3_{12}3_{34}\rangle_C$, respectively.
{\bf{b)}}: Probability densities  $\rho_{\c\c}^{(1a)}$,  $\rho_{\c\c}^{(1b)}$
and   $\rho_{\c\c}^{(0)}$
for the two heavy quarks as a function of
interquark distance
projected on the spin 1 states
$|1_{12}0_{34}\rangle_S$ and $|1_{12}1_{34}\rangle_S$
and on spin 0 state
$|0_{12}1_{34}\rangle_S$, respectively.
}
\label{slccspincolor}
\end{figure}

As in the previous subsection for the $\T_{\b\b}$ we now repeat the two-quark
probability density analysis for the $\T_{\c\c}$ system.
In Figs. \ref{slcc} and \ref{slccspincolor} the
probability densities their projections on colour and spin states as a
function of interquark distance are shown.

We can see in Fig. \ref{slcc} that the wave function between heavy quarks
is much broader and has an exponential tail at large distances. 
If we look at the structure of the quark-quark probability density 
in Fig. \ref{slccspincolor}a we see that at around $r\sim$ 1 fm sextet configurations
become larger than triplet ones and soon after the ratio of the colour
configurations stabilizes at 2. This supports the picture of molecular binding
of the $\D$ and $\D^*$ meson in the $\T_{\c\c}$.
This can also be confirmed from the results shown in Fig. \ref{slccspincolor}b
where at distances larger than 1 fm the probability for two heavy quarks
in spin 0 and spin 1 states is equal. This follows from spin recoupling
\begin{eqnarray}
|1_{13} 0_{24}\rangle_S
=&-\sqrt{\frac{1}{2}}\bigg[
|1_{12}1_{34}\rangle_S+\frac{1}{\sqrt{2}}(|1_{12}0_{34}\rangle_S
-|0_{12}1_{34}\rangle_S)\bigg],
\nonumber\\
|0_{13} 1_{24}\rangle_S
=&-\sqrt{\frac{1}{2}}\bigg[
|1_{12}1_{34}\rangle_S-\frac{1}{\sqrt{2}}(|1_{12}0_{34}\rangle_S
-|0_{12}1_{34}\rangle_S)\bigg].
\nonumber
\end{eqnarray}
%If we assume that the tetraquark spin wave function is symmetric due to
%permutation of identical particles and thus has the form
If we assume that in the ground state orbital and also colour
wave function is symmetric/antisymmetric with respect to permutation of
identical particles ($\sim |1_{13}1_{24}\rangle_C +/- |1_{14}1_{23}\rangle_C$),
the spin wave function must be antisymmetric and has the form
$$|\psi\rangle \sim 1/\sqrt{2}(|1_{13} 0_{24}\rangle_S-|0_{13}
1_{24}\rangle_S)=-1/\sqrt{2}(|1_{12}0_{34}\rangle_S-|0_{12}1_{34}\rangle_S).$$
The contribution of the $  |1_{12}0_{34}\rangle_S$ and of the
$|0_{12}1_{34}\rangle_S$
configurations is thus equal.
Similar conclusion also holds, if we recouple to $|1_{14}0_{23}\rangle _S$ and 
$|0_{14}1_{23}\rangle _S$. All relevant spin recouplings can be found in 
\cite{BS}.  

\begin{figure}[ht!]
\includegraphics[width=\linewidth]{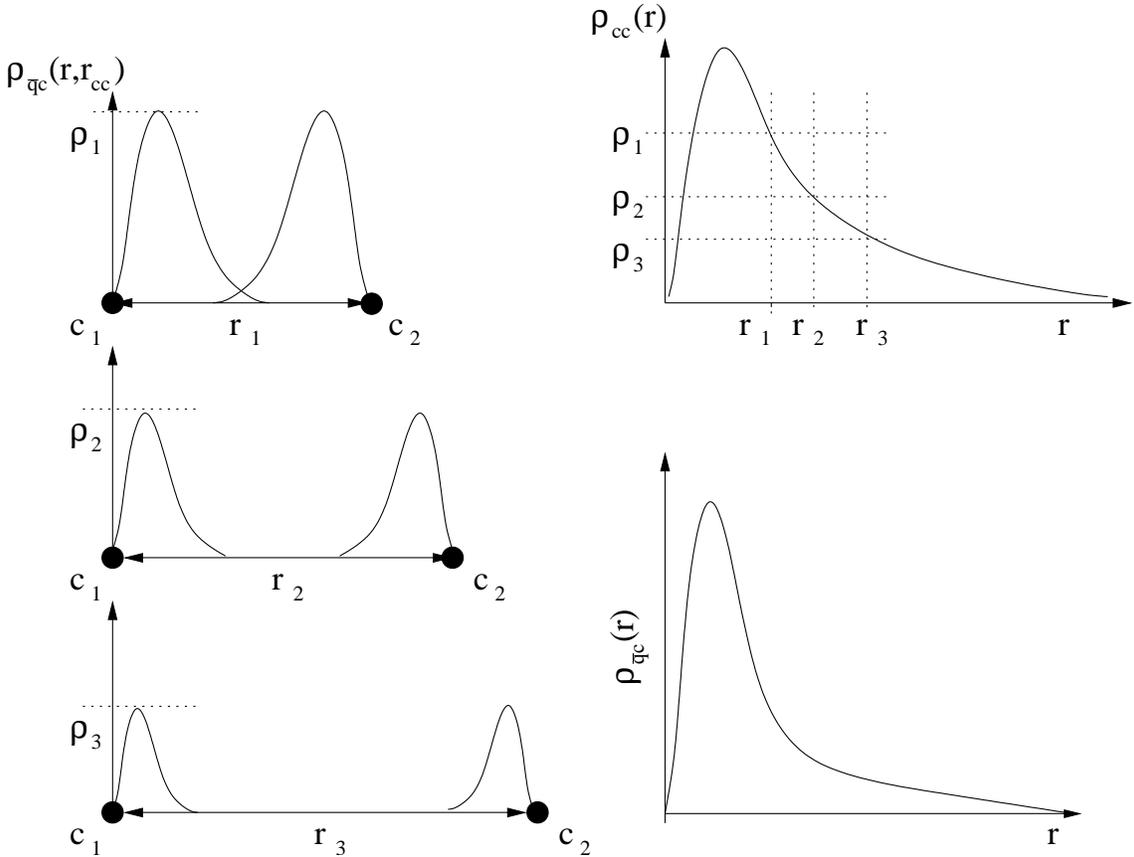}
\caption{%
Decomposition of the quark-antiquark probability density $\rho_{\bar{\q}c}$
(bottom right) into contributions
corresponding to different distances between heavy quarks (left).
 The details are explained in the text.}
\label{slsklop13}
\end{figure}

The probability density of the light antiquarks $\rho_{\bar{\q}\bar{\q}}$
shows similar behaviour, while
the radial dependence of the quark-antiquark probability density 
$\rho_{\bar{\q}\c}$ has a strong peak
at smaller distances and a twice lower tail than the $\rho_{\bar{\q}\bar{\q}}$
and $\rho_{\c\c}$ (Fig. \ref{slcc}).
How this structure appears is shown in Fig. \ref{slsklop13}. Due to
symmetrization of the configurations into which we expand the tetraquark
wave function, the first light antiquark is bound to both heavy
quarks $\c_1$ and $\c_2$ with equal probability. The probability density
$\rho_{\bar\q\c}(r,r_{\c\c})$
of finding the first light antiquark at the interquark distance $r$ from
the first heavy quark when the distance between the heavy quarks $r_{\c\c}$ is
$r_1$, $r_2$ and $r_3$ is shown on the left hand side of Fig. \ref{slsklop13}.
The heights $\rho_i$ are proportional to the probability density of finding
two heavy quarks at the interquark distance $r_i$, as is schematically depicted
on the top right of  Fig. \ref{slsklop13}. After
integrating over all possible interquark positions $r_i$
we obtain a strong peak when the light antiquark is bound to the first
heavy quark $\c_1$ and a long tail when
it is bound to the second heavy quark $\c_2$. If we ignore the interference
between these two situations which appears when $r_i$ is
smaller
than the size of the free D meson, half of the probability is in
the peak and the other half in the tail.

\begin{figure}[ht!]
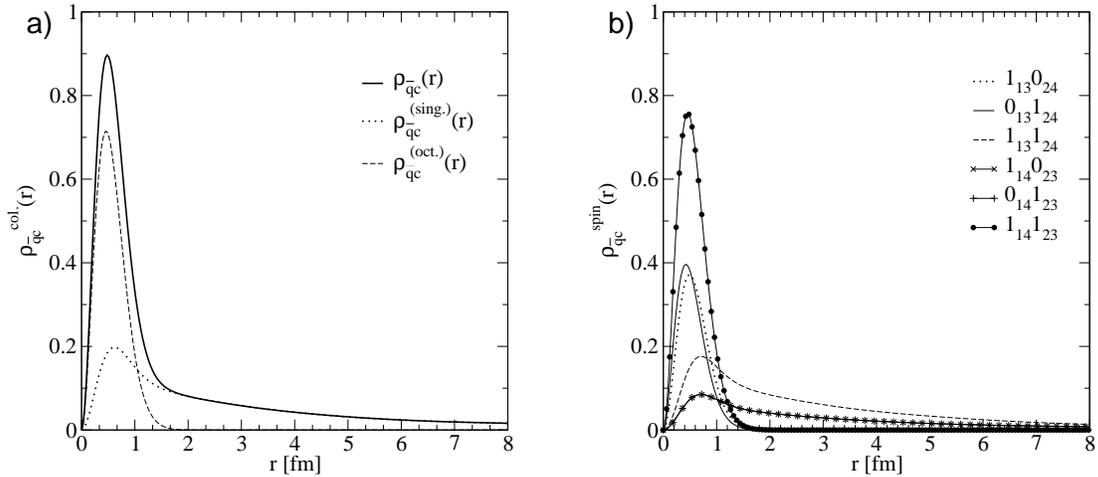

\includegraphics[width=0.45\linewidth]{Bh-qqccN140-barva.eps}
\hspace{0.05\linewidth}
\includegraphics[width=0.45\linewidth]{Bh-qqccN140-spin13.eps}
\caption {%
Results of calculations with the Bhaduri potential.
{\bf{a)}}: Probability densities  $\rho_{\bar{\q}\c}^{\mathrm{(sing.)}}$ and
$\rho_{\bar{\q}\c}^{\mathrm{(oct.)}}$
of the quark and antiquark
projected on the colour singlet
$|1_{14}1_{23}\rangle_C$ and colour octet states
$|8_{14}8_{23}\rangle_C$, respectively.
{\bf{b)}}: Projection of Probability density  $\rho_{\bar{\q}\c}$ 
on various spin states. The $|0_{13}1_{24}\rangle _S$ and
$|1_{13}0_{24}\rangle _S$  projections are almost exactly (after
renormalization) the probability densities of $\D$ and $\D^*$ mesons,
respectively.
}
\label{slccspincolor13}
\end{figure}

The colour and spin decomposition of the quark-antiquark probability 
density $\rho_{\bar{\q}c}$
is shown in Fig. \ref{slccspincolor13}. We see in Fig. \ref{slccspincolor13}b 
that at small distances ($r<1$ fm) where the peak is situated, the probability
density between first heavy quark (particle 1 from Fig.\ref{ks}) 
and first light antiquark (particle 3 from Fig.\ref{ks}) 
corresponds to the pseudoscalar 
D meson $|0_{13}1_{24}\rangle_S$
and the vector $\D^*$ meson $|1_{13}0_{24}\rangle_S$.
To understand the spin structure of the tail,
we also present in the same graph the projection on the spin
$|1_{14}0_{23}\rangle_S$,  $|0_{14}1_{23}\rangle_S$ and  $|1_{14}1_{23}\rangle_S$
states. 
We see that the dominant spin $|1_{13}1_{24}\rangle_S$ channel here can
be decomposed into the $|0_{14}1_{23}\rangle_S$ and $|1_{14}0_{23}\rangle_S$
contribution. This means that for larger separations $r$ 
%between 
%first heavy quark (particle 1) and first light antiquark (particle 3),
the heavy quark
dresses with the second light antiquark (particle 4) into the $D$ or $D^*$ meson. 
%
%We see that the dominance of the  spin $|1_{13}1_{24}\rangle_S$
%structure in the tail occurs due to the equal admixture of  $|0_{14}1_{23}\rangle_S$ and
%$|1_{14}0_{23}\rangle_S$ states. 
%
From this we can conclude that there is no
appreciable contribution from the D$^*$D$^*$ configuration 
also for large $r$.
For large $r$ on can see in Fig. \ref{slccspincolor13}a that the first
light antiquark combines with the second heavy quark into a
colour singlet state.
The octet colour dominance at small interquark distances is due to the
formation of diquark-antidiquark structure, 
similar as to that in $\T_{\b\b}$.
But in $\T_{\c\c}$ this is not a dominant structure, it represents only about a third
of the total probability in the case of the AL1 potential and even less for
the Bhaduri potential.

\newpage

\begin{figure}[ht!]
\centering
\includegraphics[width=2.2in]{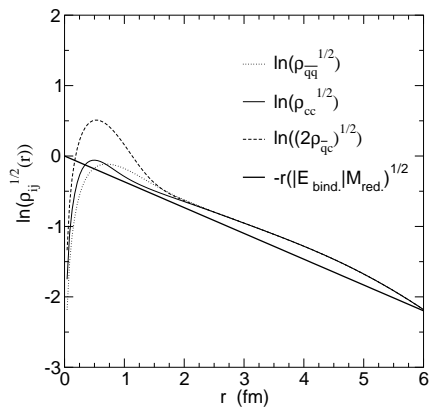}
\caption {%
Logarithms of probability density as a function of interquark distances
compared with the analytically expected slope
}
\label{slccasimp}
\end{figure}

We also analysed the decrease of the tail 
of the probability densities from Fig. \ref{slcc}a.
At large distances one can approximate the tetraquark as a bound state of
two mesons which is described by the Sch\"odinger equation
\begin{eqnarray}
-\frac{\hbar^2}{2m_r}\frac{d^2}{dr^2}\psi+V(r)\psi+E\psi,\nonumber
\end{eqnarray}
where $m_r$ is the reduced mass of $\D$ and $\D^*$.
Assuming that at large distances the colour wave function is singlet, the
potential $V(r)$ should approach zero and thus the asymptotic 
behaviour of $\psi$ should be
\begin{eqnarray}
\psi(r\to 0)\to\exp(-\kappa r),\qquad \kappa=\sqrt{|E_{\mathrm{bind.}}|
M_{\mathrm{red.}}}/\hbar c,
\label{exp1}
\end{eqnarray}
where $E_{\mathrm{bind.}}$ is the binding energy of the tetraquark
(2.7 MeV for AL1) and $M_{\mathrm{red.}}$ the reduced mass of the
$\D$ and $\D^*$ meson.
This can be very clearly seen in  Fig. \ref{slccasimp} where we plotted
the logarithm of the wave functions. The quark-antiquark probability
density is multiplied by 2 as explained in the previous paragraph.
All three interquark probability densities follow the predicted
exponential decrease from Eq. \ref{exp1}.

\section{Three-body interaction}

The idea of introducing a small amount of three-body interaction into
nonrelativistic potential models is not new.
In the baryon sector this additional interaction can be used to
better reproduce the ground state spectroscopy \cite{al1}.
But generally, a slight three-body interaction cannot produce
any significant changes in other baryon properties,
and the desirable shift of levels can also be reproduced by
modification of the parameters in the two-body potential.

Before we introduce such an interaction into the tetraquark
system we shortly discuss the structure of this interaction.
For the radial part we take the simplest possible radial dependence --
the smeared delta function of the coordinates of the three interacting
particles. Since we are working with the colour$\cdot$colour type
potential it is natural that also the three-body potential possesses
some colour structure. The colour factor in the two-body
Bhaduri or AL1 potential is proportional to the first (quadratic)
Casimir operator C$^{(1)}$; C$^{(1)}=\lambda\cdot\lambda$. Following this, we
introduce in the three-body potential the second C$^{(2)}$ (cubic) Casimir
operator C$^{(2)}= d^{abc}\lambda_a\cdot\lambda_b\cdot\lambda_c$.
A deeper discussion of the properties that the colour dependent three-body
interaction must fulfil can be found in \cite{dma}, \cite{dma2}, \cite{s2}.

It should be noted that in the baryon sector such a colour structure is
irrelevant since there is only one colour singlet state and thus the colour
factor is just a constant which can be included into the strength of the
potential. In tetraquarks the situation is different since there are
two colour singlet states: $\bar3_{12}3_{34}$ and  $6_{12}\bar6_{34}$
(or $1_{13}1_{24}$ and $8_{13}8_{24}$ after recoupling). The
three-body force operates differently on these two states and one can
anticipate that in the case of the weak binding it can
produce large changes in the structure of the tetraquark.
This cannot be otherwise produced simply by reparameterization
of the two-body potential, so the weakly bound tetraquarks
are a very important laboratory for studying the effect of such
an interaction.

 The form of the three-body interaction we introduced into
 the tetraquark is
\begin{eqnarray}
V^{3{\mathrm{b}}}_{qq\bar q}(\vec{r}_i,\vec{r}_j,\vec{r}_k)&=&-\frac{1}{8}d^{abc}
\lambda_i^a
\lambda_j^b \lambda_k^{c*} U_0 \exp[-(r_{ij}^2+r_{jk}^2+r_{ki}^2)/r_0^2],
\nonumber\\
V^{3{\mathrm{b}}}_{q\bar q\bar q}(\vec{r}_i,\vec{r}_j,\vec{r}_k)&=&\frac{1}{8}d^{abc}
\lambda_i^a\lambda_j^{b*} \lambda_k^{c*} U_0
\exp[-(r_{ij}^2+r_{jk}^2+r_{ki}^2)/r_0^2].\nonumber
\end{eqnarray}
Here \textit{r}$_{ij}$ is the distance between i-th and j-th (anti)quark,
 and similarly for
\textit{r}$_{jk}$ and \textit{r}$_{ki}$.
$\lambda_a$ are the Gell-Mann colour matrices and $d^{abc}$ are the SU(3)
structure constants
($\{\lambda^a,\lambda^b\}=2d^{abc}\lambda^c$).

\begin{figure}[h h h]
\includegraphics[width=0.43\linewidth]{tridelcna1.eps}
\hspace{0.05\linewidth}
\includegraphics[width=0.47\linewidth]{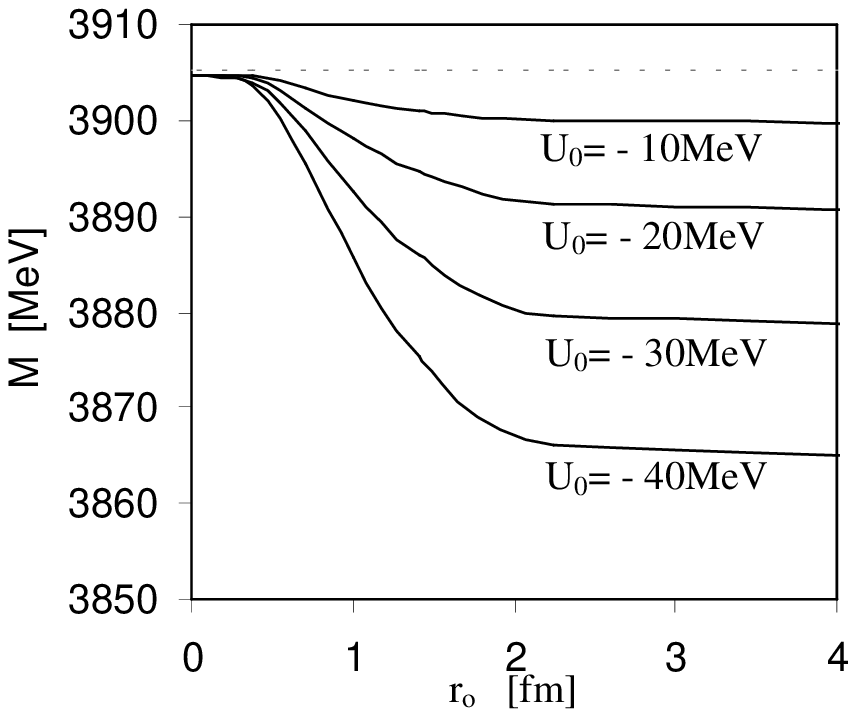}
\put(-180,135){b)}
\put(-370,135){a)}
\caption{%
{\bf{a)}}: Probability density between two c quarks in the $\T_{\c\c}$
tetraquark as
a function of interquark distance
for three different values of the strength of the three-body potential.
{\bf{b)}}: Mass of the $\T_{\c\c}$ tetraquark as a function of
the smearing parameter $r_0$ in the three-body interaction for four different
strengths $U_0$ of the three-body potential.
}
\label{sl3body}
\end{figure}
The diagonal matrix elements of the colour part of the three-body interaction
between two quarks and an antiquark are -5/18 and 5/9 for
$|\bar{3}_{12} 3_{34} \rangle$ and $|6_{12}\bar{6}_{34} \rangle$ colour
states, respectively. If the strength of this interaction $U_0$ is negative
it will lower the states with diquark-antidiquark configuration and
increase the binding as can be seen
in Fig. \ref{sl3body}b. For a strong three-body interaction $\T_{\c\c}$
loses the molecular structure and the triplet-triplet colour configurations
become dominant and the $\T_{\c\c}$ tetraquark becomes similar to
$\T_{\b\b}$. A drastic change in the width of the probability density
can already be seen for strength $U_0$ = -20 MeV. In the baryon sector
such an interaction would merely lower the states by about $U_0$
so it would have no dramatic effect nor would it
spoil the fit to experimental data. 
Since the predicted energies of ground state baryons for the
Bhaduri and AL1 potential are above the experimental values, this is
actually a desirable feature.
The dependence of the mass of the $\T_{\c\c}$ tetraquark on the
strength of the potential $U_0$ and on the smearing of this potential
is shown in Fig. \ref {sl3body}b. One can see that for large smearing
the mass of the tetraquark is shifted by about $4\frac{5}{18}U_0$ in agreement
with the dominance of the triplet-triplet colour configuration in this state.

\section{Production and detection of $\T_{\c\c}$}
We shall focus only on the possibility of detecting the $\T_{\c\c}$
tetraquark, since even at LHC the production of the $\T_{\b\b}$ tetraquark
would be below the rate where one could hope to detect it.

Since double charmed baryons were probably detected
at SELEX \cite{sx} one can expect that if the $\T_{\c\c}$ is bound,
it was also produced there, with a production rate about
ten times smaller than double charmed baryons.
This estimate is based on the fact that
a heavy quark gets dressed with a light antiquark into a heavy meson
with a probability of
roughly 0.9, and with a probability of 0.1 it combines with two light
quarks into a $\Lambda$ baryon \cite{dress}. But since SELEX found,
with their cuts, only fifty candidates for
double charmed baryons, the statistics for detecting double charmed
tetraquarks should be improved. Another experiment
where one might look for  $\T_{\c\c}$ could be provided by
LHC where one can estimate the production rate  as large as
$10^4$ events/hour \cite{RJ}.

However, the $\T_{\c\c}$ tetraquark has  a molecular structure
in which the mesonic wave function
is not strongly influenced by the other meson. The large mean square
radius of the $\T_{\c\c}$ tetraquark has  consequences for the
production mechanism. It is no longer required that the two
c quarks come close  to produce first a diquark which then
gets dressed by light antiquarks.
This can give a larger production rate at SELEX than estimated above.
This also makes machines like the RHIC ion collider \cite{rhic}
interesting candidates for searching for tetraquarks.

There is also an interesting possibility of
production and detection of the $\T_{\c\c}$ tetraquark in 
B-factories.  Belle \cite{cccc} has reported a measurement of
double charm production in $\mathrm{e}^+\mathrm{e}^-$ 
annihilation at $\sqrt{s}$ = 10.6 GeV
and found that
$$\sigma(\mathrm{e}^+\mathrm{e}^-\to J/\psi \c\bar{\c})\sim \mathrm{1pb,}$$ 
which corresponds to about 2000 events.
Since the total mass of four D mesons is close to the
c.m. energy, the c quarks created in this process have
small relative momentum which is very important in
the $\T_{\c\c}$ production.

The main problem with detection of the weakly bound
$\T_{\c\c}$ tetraquark is how to distinguish the pion or photon
emitted by the decay of the free $\D ^*$ meson from the one
emitted by the  $\D ^*$ meson bound inside the tetraquark.
We can exploit the fact that the phase
space for $\D ^*\to\D+\pi$ decay is very small.
%(see Table \ref{Dmezoni}).
Note that $m_{\rm D^{*+}}-m_{\rm D^+}-m_{\pi^0}= 5.6 \pm 0.1\,{\rm MeV},\quad
m_{\rm D^{*0}}-m_{\rm D^0}-m_{\pi^0}=7.1 \pm 0.1\,{\rm MeV},\quad
m_{\rm D^{*+}}-m_{\rm D^0}-m_{\pi^+}= 5.87 \pm 0.02\,{\rm MeV}.$ 
This has a strong impact on the branching ratio
between radiative and hadronic decay. Since the  $\D ^*$ meson
inside the tetraquark is not significantly influenced by the
other D meson (Fig. \ref{slsklop13}) in the tetraquark,
we expect that the partial width for the magnetic dipole
M1 transition would be very close to the width of the free meson
while the width for hadronic $\D ^*\to\D+\pi$ decay will decrease with stronger
binding and will become energetically forbidden below the $\D +\pi$ threshold.
The hadronic decay of the $\T_{\c\c}$ tetraquark is a three-body
decay which is commonly represented by the Dalitz plot.

\begin{figure}[here]
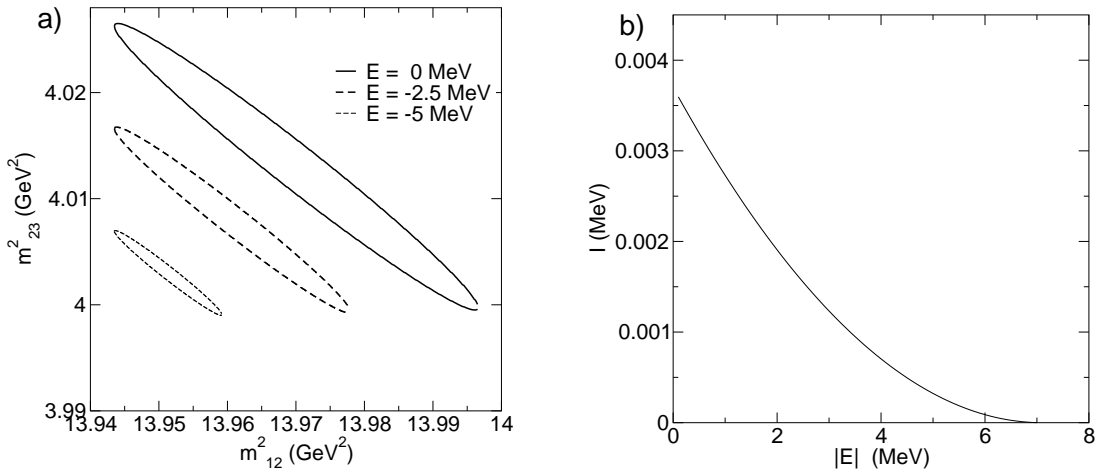

\includegraphics[width=0.45\linewidth]{dalitz1.eps}
\hspace{0.05\linewidth}
\includegraphics[width=0.45\linewidth]{dalitz2.eps}
%\resizebox{2.15in}{2.15in}{\includegraphics[width=3in]{vfBBn80color.eps}}
\caption {%
{\bf{a)}}: Dalitz plot, {\bf{b)}}: Integrated Dalitz plot.
}
\label{slab}
\end{figure}

Let us assume that the $\T_{\c\c}$ tetraquark is below
the $\D+\D^*$ threshold but above the $\D+\D+\gamma$ and $\D+\D+\pi$
as was the case in our nonrelativistic potential models.
Then
the partial decay rate for the $\T_{\c\c}\to$D+D+$\pi$ is given by
\begin{equation}
d\Gamma=\frac{1}{(2\pi)^3}\frac{1}{32M^3}\overline{|{\cal M}|^2}
dm_{12}^2dm_{23}^2
\end{equation}
where particles 1 and 2 are two final $\D$ mesons and particle 3 
is a $\pi$ emerging from the decaying tetraquark.
Here  $m_{12}^2=(p_D+p_D)^2$ and $m_{23}^2=(p_D+p_\pi)^2$
and M is the mass of the tetraquark.
Since the
total masses of the $\D ^*+D$ and $2\D+\pi$ are so close there is a strong
isospin violation in the decay which cannot be reproduced with
the Bhaduri or AL1 potential where the $\D ^*$ and the $\D$ isospin doublets
are degenerate.
We shall not try to modify the interaction to accommodate the dependence of the
decay on the isospin of the particles, but we shall rather
work with the experimental masses 
taken from the PDG \cite{pdg}.
The allowed region of integration over $dm_{12}^2$ and $dm_{23}^2$
for three different binding energies is plotted
in Fig. \ref{slab}a. If we assume $\overline{|{\cal M}|^2}$ is constant, which is very
plausible in our case, the allowed
region will be uniformly populated with experimental events so that the
measured partial decay rate $\Gamma$ will be proportional to the kinematically
allowed area from Fig. \ref{slab}a. The dependence of this area $I$
as a function of the binding energy of the tetraquark
is shown in Fig. \ref{slab}b.

Up to now, we have discussed only true bound states,
but the $\T_{\c\c}$ tetraquark
can also be a resonant state above the $\D +\D^*$ threshold. Then if the
resonance is situated near the threshold, beside the
$\T_{\c\c}\to \D +\D^*$ decay there will still be a significant
fraction of hadronic $\T_{\c\c}\to \D +\D +\pi$ decays. This region of
positive binding energy is not presented in Fig. \ref{slab}.
 But one can see that in a similar manner as in the case
of weakly bound tetraquarks the low-lying resonant state can 
be identified from the Dalitz plot.

\begin{table}[here]
\caption{Mean distance between two heavy quarks
$\langle r_{\c\c}\rangle$ in the $\T_{\c\c}$
tetraquark and the value of the centrifugal potential for L=2 state
for the Bhaduri and the AL1 potential.}
\label{tabelaR}
\vspace {0.5cm}
\begin{tabular}{c c c}
\hline
 & $\langle r_{\c\c} \rangle$ 
 &$(\hbar^2L(L+1)/2M_{red})\langle 1/r_{\c\c}^2\rangle$\\
\hline
Bhaduri (L=2) & 2.4 fm   &  174 MeV \\
AL1 (L=2)&   1.6 fm &  232 MeV \\
\hline
\end{tabular}
\end{table}

As a remark, we present an alternative way of detecting the
weakly bound tetraquark.
Since the mean radius of the tetraquark is  large, as one can see from
Table \ref{tabelaR}, the
centrifugal barrier for the L=0 $\to$ L=2 transition is comparable with
the available energy in the $\D ^*$ decay, so there is also a possibility
of the electric quadrupole transition E2. %This decay can go through two
%different mechanisms as shown in Fig. \ref{sle2}.
%While the first one
%(Fig. \ref{sle2} a)) can be neglected
%due to the very small admixture of a $\D ^*\D ^*$
%state in the  $\T_{\c\c}$ tetraquark, the second one
%(Fig. {sle2} b)) can still
%give rise to the measurable effect.
This is a two pion exchange process
which is beyond the scope of the potential model used here.

%\begin{figure}[here]
%\centering
%\includegraphics[width=1.2in]{e2prehod4.eps}
%\caption {E2transition2 }
%\label{piex}
%\end{figure}

\section{Conclusions}

We have shown that in popular nonrelativistic potential models
the $\T_{\c\c}$ tetraquark is bound against the $\D\D^*$ threshold
and that it has  a {\em molecular} structure. 
Therefore the approximation based on the
assumption of the {\em atomic} $\bar{\Lambda}_\b$-like structure
which suggests that the system is not bound \cite{JR}
is not valid. This dramatically
different situation as compared with the $\T_{\b\b}$ tetraquark makes the
$\T_{\c\c}$ tetraquark an interesting laboratory for more profound
studies of the nature of the interactions between quarks, 
since such a weakly bound
system is more sensitive to the detailed features of the interaction.

As a signature for the $\T_{\c\c}=\D\D^*$ tetraquark one might exploit the
very small phase space of the $\D*\to\D\pi$ decay which is very
sensitive to the binding energy of $\D^*$ to $\D$.

It seems tempting to compare the $DD^*$ dimeson with the recently
found $D\bar{D^*}(3.872)$ state which is just above the $D\bar{D}^*$
threshold. This is, However, a completely different situation due to the
low $J/\psi\,\pi$ or $J/\psi\,\eta$ thresholds, which
are degenerate for both potentials used here.
Thus a delicate coupled channel calculation
would be needed in the search for resonances which 
is beyond the scope of this paper. Our energy minimisation
procedure gave in fact the threshold energy 3232 MeV
for spin 1 $c\bar cu\bar d$ state calculated with the 
Bhaduri potential, consistent with ref.\cite{SB}, where due to their
basis they obtained somewhat higher value of 3468 MeV.

We did not study in detail the $\c\c\bar{\s}\bar{\u}$ tetraquark since
it has a lower threshold, $\D_s\D$ rather than $\D_s\D^*$ or $\D_s^*\D$,
and it is not likely to be bound.

\paragraph{Acknowledgement.}
The authors would like to acknowledge the encouragement by Jean-Marc
Richard to look whether or not the $\D\D^*$ system might be bound after all.
Acknowledgment is also due to the collaboration with Danielle Treleani
and Alessio Del Fabbro on questions of production of the $\T_{\b\b}$
and $\T_{\c\c}$ tetraquarks.

This work was supported by the Ministry of Education, Science and Sport
of the Republic of Slovenia.

%\section{Appendix}
%\subsection{Configurations}
\appendix
\section{ Configurations}
\begin{figure}[here]
\centering
\includegraphics[width=4in]{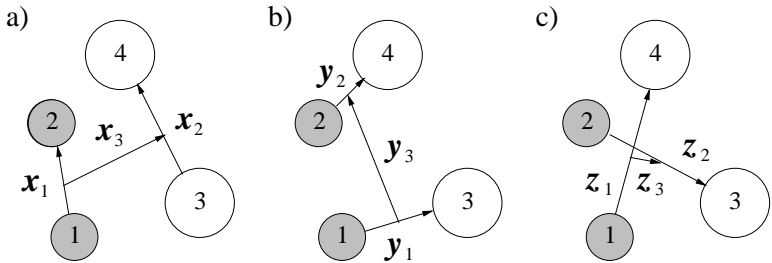}
\caption{%
Two quarks (dashed circles) and two antiquarks (empty
circles) in three different relative coordinate systems.
The orbital wave function is then a Gaussian function of
relative coordinates.
a) diquark-antidiquark: 
$K_1(C_{ij})=\exp(-\mathbf{x}_iC_{ij}\mathbf{x}_j)$,
b) direct channel: 
$K_2(C_{ij})=\exp(-\mathbf{y}_iC_{ij}\mathbf{y}_j)$,
c) exchange channel:
$K_3(C_{ij})=\exp(-\mathbf{z}_iC_{ij}\mathbf{z}_j)$.
}
\label{ks}
\end{figure}

We express the orbital part of the tetraquark wave function 
in terms of Gaussians. The coordinate systems used here
are shown in Fig. \ref{ks}. The transformation between 
various coordinate systems and some details about the
calculation of the kinetic and potential matrix elements \
are given in \cite{BS}.  

The most general form for the
ground state of the tetraquark (L=0) can be expanded
in any of the three configurations given in Fig. \ref{ks} 
\begin{eqnarray}
R=\sum_n C_n K_r(C_{ij}^{n}),\qquad C_{ij}^{n}=C_{ji}^{n},
\qquad r=1\,{\mathrm{or}}\,2 \,{\mathrm{or}}\, 3.\nonumber
\end{eqnarray}
So far it looks as if not all coordinate systems were needed,
but in numerical calculation it is convenient to 
limit the test functions to those in which
$C_{ij}=0$ if $i\ne j$. With this we reduce our
problem of optimization of 6 parameters $C_{ij}$
into three optimizations ($r=1,2,3$) of 3 parameters
 $C_{ii}=\{c_1,c_2,c_3\}$.
  Though this somewhat restricts our Hilbert 
 space we still expect that it would not have
 any significant effect on the calculation of the ground state,
 because by using all three configurations from Fig. \ref{ks}
 we can have nonzero relative angular momentum between
 two quarks $l_{12}$ or two antiquarks $l_{34}$  by using the
 systems b) and c) but still
 keep the total angular momentum zero.  
 The orbital part of the wave function has thus
 the form
\begin{eqnarray}
 R=\sum_n C_n K_r(c_1^n,c_2^n,c_3^n),
\qquad r=1,2,3.\nonumber
\end{eqnarray}

The tetraquark wave function must posses correct
symmetry against the permutation of the two quarks
$P(1,2)$ or antiquarks $P(3,4)$. The effect of 
these permutations on relative coordinates are\\
\begin{tabular}{l l }
&\\
$P(1,2)\big[K_1(c_i)\big]= K_1(c_i)$, & $P(3,4)\big[K_1(c_i)\big]= K_1(c_i)$,\\
$P(1,2)\big[K_2(c_1,c_2,c_3) \big]= K_3(c_2,c_1,c_3)$, &
$P(3,4)\big[K_2(c_i) \big]= K_3(c_i)$, \\
$P(1,2)\big[K_3(c_1,c_2,c_3) \big]= K_2(c_2,c_1,c_3)$,\vspace{1mm}&
$P(3,4)\big[K_3(c_i) \big]= K_2(c_i)$.\\
&\\
\end{tabular}\\
We see that if $c_1\ne c_2$ the configuration $K_2$ and $K_3$
loose symmetry properties with respect to permutations of
the heavy quarks. 
To obtain functions with good permutation symmetry we 
make linear combinations of configurations.
For shorter notation we introduce
$K_i(c_2,c_1,c_3)=\tilde K_i(c_1,c_2,c_3)$\\
\begin{tabular}{@{}c @{}l}
&\\
$R_1(c_i)=$ & $K_1(c_i)$,\\
\vspace{1mm}
$R_2(c_i)=$ & $K_2(c_i)+K_3(c_i)+\tilde K_2(c_i)+\tilde K_3(c_i)$,\\
\vspace{1mm}
$R_3(c_i)=$ & $K_2(c_i)-K_3(c_i)+\tilde K_2(c_i)-\tilde K_3(c_i)$,\\
\vspace{1mm}
$R_4(c_i)=$ & $K_2(c_i)-K_3(c_i)-\tilde K_2(c_i)+\tilde K_3(c_i)$,\\
\vspace{1mm}
$R_5(c_i)=$ & $K_2(c_i)+K_3(c_i)-\tilde K_2(c_i)-\tilde K_3(c_i)$.\\
&\\
\end{tabular}\\
The effect of permutation of identical (anti)quarks is then\\
\begin{tabular}{@{}c @{}r @{}r}
& &\\
$P(12)\big[R_1(c_i)\big]=$ & $P(34)\big[R_1(c_i)\big]=$ & $R_1(c_i)$,
\vspace{1mm}\\
$P(12)\big[R_2(c_i)\big]=$ & $P(34)\big[R_2(c_i)\big]=$ & $R_2(c_i)$,
\vspace{1mm}\\
$P(12)\big[R_3(c_i)\big]=$ & $P(34)\big[R_3(c_i)\big]=$ & $-R_3(c_i)$,
\vspace{1mm}\\
$P(12)\big[R_4(c_i)\big]=$ & $-P(34)\big[R_4(c_i)\big]=$ & $R_4(c_i)$,
\vspace{1mm}\\
$P(12)\big[R_5(c_i)\big]=$ & $-P(34)\big[R_5(c_i)\big]=$ & $-R_5(c_i)$.
\\
& &\\
\end{tabular}

In the spin space of four quarks we have three different spin 1 
representations. Most suitable basis for studying permutation
properties is obtained by coupling the quarks into a diquark
and antiquarks into a antidiquark. The three basis states are then 
\begin{eqnarray}
|1_{12},0_{34}\rangle,\quad|0_{12},1_{34}\rangle,\quad
|1_{12},1_{34}\rangle,
\label{eqspin}
\end{eqnarray}
and the permutation of identical particles on this states gives\\
\begin{tabular}{l l}
&\\
$P(12)[|1_{12},0_{34}\rangle]=|1_{12},0_{34}\rangle$,&
$P(34)[|1_{12},0_{34}\rangle]=-|1_{12},0_{34}\rangle$,\\
$P(12)[|0_{12},1_{34}\rangle]=-|0_{12},1_{34}\rangle$,&
$P(34)[|0_{12},1_{34}\rangle]=|0_{12},1_{34}\rangle$,\\
$P(12)[|1_{12},1_{34}\rangle]=|1_{12},1_{34}\rangle$,&
$P(34)[|1_{12},1_{34}\rangle]=|1_{12},1_{34}\rangle$.\\
&\\
\end{tabular}

In the colour space there are two different colour singlet
representations which in the diquark-antidiquark basis can
be expressed as
\begin{equation}
|\bar 3_{12},3_{34}\rangle,\quad|6_{12}\bar, 6_{34}\rangle,
\label{eqcolor}
\end{equation}
and we have\\
\begin{tabular}{l l}
&\\
$P(12)[|\bar3_{12},3_{34}\rangle]=-|\bar 3_{12},3_{34}\rangle$,&
$P(34)[|\bar3_{12},3_{34}\rangle]=-|\bar 3_{12},3_{34}\rangle$,\\
$P(12)[|6_{12},\bar 6_{34}\rangle]=|6_{12},\bar 6_{34}\rangle$,&
$P(34)[|6_{12},\bar 6_{34}\rangle]=|6_{12},\bar 6_{34}\rangle$.\\
&\\
\end{tabular}

From the original three spatial configuration shown in Fig. \ref{ks}, 
the three spin configurations given in Eq.(\ref{eqspin})  
and two colour configuration of
Eq.(\ref{eqcolor}) one can build up $3\cdot 3\cdot 2 = 18$
function. Eight of them are antisymmetric with respect to exchange of
heavy quarks and symmetric (isospin 0) with respect to
exchange of light antiquarks.\\ 
\begin{tabular}{l l}
&\\
$\psi_1=R_1(c_i)|\bar 3_{12}3_{34}\rangle_C|1_{12}0_{34}\rangle_S$,&
$\psi_2=R_1(c_i)|6_{12}\bar 6_{34}\rangle_C|0_{12}1_{34}\rangle_S$,
\vspace{1mm}\\
$\psi_3=R_2(c_i)|\bar 3_{12}3_{34}\rangle_C|1_{12}0_{34}\rangle_S$,&
$\psi_4=R_2(c_i)|6_{12}\bar 6_{34}\rangle_C|0_{12}1_{34}\rangle_S$,
\vspace{1mm}\\
$\psi_5=R_3(c_i)|6_{12}\bar 6_{34}\rangle_C|1_{12}0_{34}\rangle_S$,&
$\psi_6=R_3(c_i)|\bar 3_{12}3_{34}\rangle_C|0_{12}1_{34}\rangle_S$,
\vspace{1mm}\\
$\psi_7=R_4(c_i)|\bar 3_{12}3_{34}\rangle_C|1_{12}1_{34}\rangle_S$,&
$\psi_8=R_5(c_i)|6_{12}\bar 6_{34}\rangle_C|1_{12}1_{34}\rangle_S$.
\\
&\\
\end{tabular}\\
For better description of weakly bound states we also add
additional configurations which cannot be decomposed into
a simple product of orbital, colour and spin parts.\\
\begin{tabular}{l}
\\
$\psi_9=\bigg(\big(K_2(c_i)+\tilde K_2(c_i)\big)|1_{13}1_{24}\rangle_C+
\big(K_3(c_i)+\tilde K_3(c_i)\big)|1_{14}1_{23}\rangle_C
\bigg)|0_{12}1_{34}\rangle_S$,
\vspace{1mm}\\
$\psi_{10}=\bigg(\big(K_2(c_i)+\tilde K_2(c_i)\big)|1_{13}1_{24}\rangle_C-
\big(K_3(c_i)+\tilde K_3(c_i)\big)|1_{14}1_{23}\rangle_C
\bigg)|1_{12}0_{34}\rangle_S$,
\vspace{1mm}\\
$\psi_{11}=\big(K_2(c_i)-\tilde K_2(c_i)\big)|1_{13}1_{24}\rangle_C|0_{13}1_{24}\rangle_S+
           \big(K_3(c_i)-\tilde
K_3(c_i)\big)|1_{14}1_{23}\rangle_C|0_{14}1_{24}\rangle_S$,
\vspace{1mm}\\
$\psi_{12}=\big(K_2(c_i)-\tilde K_2(c_i)\big)|1_{13}1_{24}\rangle_C|1_{13}0_{24}\rangle_S+
           \big(K_3(c_i)-\tilde
K_3(c_i)\big)|1_{14}1_{23}\rangle_C|1_{14}0_{23}\rangle_S$,
\vspace{1mm}\\
$\psi_{13}=\big(K_2(c_i)-\tilde K_2(c_i)\big)|1_{13}1_{24}\rangle_C|1_{13}1_{24}\rangle_S+
           \big(K_3(c_i)-\tilde K_3(c_i)\big)|1_{14}1_{23}\rangle_C|1_{14}1_{23}\rangle_S$.
\\
\\
\end{tabular}\\
It is obvious that these configurations also respect permutation symmetry.

If we have a strong quark mass asymmetry we expect 
clustering of heavy quarks into a diquark which then results in the 
{\it{atomic}} structure of the tetraquark
\cite {BS},
so that the first coordinate system in Fig. \ref {ks} is more suitable and the
 dominant colour configuration has the
 diquark in antitriplet and the antidiquark in triplet colour state.
On the other hand, if the binding is weak, the direct and exchange meson-meson
channels are more adequate and the important configuration has  singlet-singlet
colour structure.

\section{Potential models}
For solving Schr\"odinger equation in non-diagonal basis we
use general diagonalization of the Hamiltonian
\begin{eqnarray}
\langle \psi_k^m | H|\psi_l^n\rangle = \langle \psi_k^m | W_k|\psi_l^n\rangle +
 \langle \psi_k^m | \sum_{^{i=1..3;} _{j=i+1..4}}V_{ij}|\psi_l^n\rangle,\nonumber\\
 k,l = 1,..13;\qquad m,n = 1,..N_{max};\nonumber
\end{eqnarray}
where $|\psi_k^m\rangle$ is the m-th basis function (See Appendix D) and
$N_{max}$ is the dimension of the basis. 
The kinetic energy operator written in the basis a) of Fig.\ref{ks}
has the form
%\begin{displaymath}
\begin{eqnarray}
 \langle \psi_k^m | W_k|\psi_l^n\rangle =-6\langle \psi_k^m |\psi_l^n\rangle
{\mathrm{Tr}}\Big[ (C^m+C^n)^{-1}C^m T C^n \Big],\nonumber
\end{eqnarray}
\begin{displaymath}
T=\frac{\hbar^2 c^2}{2}\left( \begin{array}{c c c}
\frac{m_1+m_2}{2m_1 m_2} & 0 & \frac{m_2-m_1}{2\sqrt{2}m_1 m_2} \\
0 & \frac{m_3+m_4}{2m_3 m_4} & \frac{m_3-m_4}{2\sqrt{2}m_3 m_4} \\
\frac{m_2-m_1}{2\sqrt{2}m_1 m_2} & \frac{m_3-m_4}{2\sqrt{2}m_3 m_4} &
 \frac{1}{4}\sum_{i=1}^4\frac{1}{m_i} \end{array}\right).
\end{displaymath}
We tested two different potentials. First one was proposed by
Bhaduri and collaborators \cite{Bh} and the improved one proposed by
Silvestre-Brac and Semay \cite{al1} which we will denoted by AL1 potential.
\begin{itemize}
\item[-]Bhaduri potential:
\begin{eqnarray}
V^{B}_{ij}=-\frac{\lambda^{C}_i}{2}\cdot \frac{\lambda^{C}_j}{2}
\Bigg( U_0+\frac{\alpha}{r_{ij}}+\beta r_{ij}+
\alpha\frac{\hbar^2}{m_i m_j c^2}\frac{e^{-r_{ij}/r_0}}{r_0^2r_{ij}}
\sigma_i\cdot \sigma_j\Bigg),\nonumber\\
r_{ij}=|\vec r_i-\vec r_j|;\nonumber
\end{eqnarray}
\begin{tabular}{l l}
$m_\b=5259$ MeV,& $m_\c=1870$ MeV,\\
$m_\s=600$ MeV,& $m_\u=m_d=337$ MeV,\\
$U_0=685$ MeV,&$\alpha=77$ MeVfm,\\
$\beta=706.95$ MeV/fm,& $r_0=0.4545$ fm.\\
\end{tabular}

\item[-]AL1 potential
\begin{eqnarray}
V^{AL1}_{ij}=-\frac{\lambda^{C}_i}{2}\cdot \frac{\lambda^{C}_j}{2}
\Bigg( U_0+\frac{\alpha}{r_{ij}}+\beta r_{ij}+
\tilde\alpha\frac{2\pi\hbar^2}{3 m_i m_j c^2}\frac{e^{-r_{ij}^2/r_0^2}}
{\pi^{3/2}r_0^3}\sigma_i\cdot \sigma_j\Bigg),\nonumber\\
r_0(m_i,m_j)=A\Bigg(\frac{m_i+m_j}{2m_im_j}\Bigg)^B,\qquad
r_{ij}=|\vec r_i-\vec r_j|;\nonumber
\end{eqnarray}
\begin{tabular}{l l l}
$m_\b=5227$ MeV,& $m_\c=1836$ MeV,& $A=1.6553$ GeV$^{B-1}$,\\
$m_\s=577$ MeV,& $m_\u=m_d=315$ MeV,& $B=0.2204$,\\
$U_0=624.075$ MeV,&$\alpha=74.895$ MeVfm,&\\
$\beta=629.315$ MeV/fm,&$\tilde\alpha=274.948$ MeVfm.&\\
\end{tabular}
\end{itemize}

\section{ Effective potential}

\begin{figure}[ht!]
\centering
\includegraphics[width=3in]{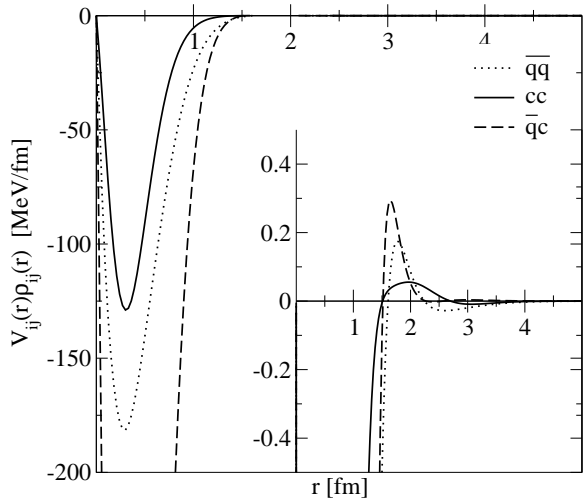}
\caption{%
Potential densities $v_{ij}$ between (anti)quarks as calculated from
Eq. \ref{vdw} for Bhaduri potential. Inserted : Enlarged section of the
figure, where
van der Waals attraction and 
medium-range repulsion can be  seen.
}
\label{efpot}
\end{figure}

Since the $T_{cc}$ system is very weakly bound, a comment 
on long-range colour van der Waals forces is in order 
\cite{isgur83,isgur83-11a, isgur83-11b}. This forces appears
due to the colour structure of the confining potential.
They act between colour singlet clusters and their asymptotic
behaviour depends on the  power $d$ of the long range
potential. In the case of the linear confining interaction
we have
$$
V(r)_{v.d.Waals}=\mathcal{O}(r^{d-4})=\mathcal{O}(r^{-3})
$$
This interaction appears due to the colour polarisation 
of two mesons in the colour singlet state. 
Such a long-range colour force is not
physically allowed and is artefact of the potential approach.
It is not present in the full QCD where quark-anriquark pair
creation from the confining field energy would produce an
exponential cut-off of this residual interaction. One might be 
concerned, however, that this spurious interaction could 
have some misleading effects, when the system is very close to the
threshold, as is the case with the $T_{cc}$ tetraquark. 
To show that this is not the case, we present in Fig. \ref{efpot}
the effective potential densities
\begin{eqnarray}
v_{ij}(r)=\langle\psi|V_{ij}(r_ij)\delta(r-r_ij)|\psi\rangle=
V_{ij}(r)\rho_{ij}(r).
\label{vdw}
\end{eqnarray}

In Fig. \ref{efpot} one can see that this effect is indeed
present at large separations ($r>2$ fm) but it is extremely small.
Integrating this attractive tail of the potential
contributes less than 100 keV to the binding of the system.
Another interesting feature of the 
effective potential shown in Fig.  \ref{efpot} is the 
repulsion between quarks at the medium distance between
quarks (1.5 fm$>r>$ 2 fm). The maximal value of
potential barrier is 
$V_{ij}(r\sim 1.5\,{\mathrm{fm}})=v_{ij}/\rho_{ij}=1$ MeV, too small
to produce an additional resonant state.

At even shorter distances we have a strong 
attractive interaction between (anti)quark-(anti)quark pairs,
in particular there is a strong attraction between
quark and antiquark. The major part of this interaction 
bind them into the $D$
or $D^*$ meson, since the molecular structure is dominant in the
$T_{cc}$ tetraquark. The residual part of this interaction helps together with  
the forces between quark-quark and antiquark-antiquark pairs to bind the
two mesons into the tetraquark.
This interactions are effective at small interquark distances ($r<1$ fm),
where the atomic configuration is important. Therefore it is crucial
even for the tetraquarks with molecular structure
that the model used in the calculation is capable of describing
accurately also the baryon spectra. 

\section{ Numerics}

We solve our four body problem by diagonalization of the Hamiltonian
in a space spanned by Gaussian function.
We built our basis step by step so that at each step all
configurations $|\psi_\alpha\rangle$ ($\alpha=1,...13$) are tested and 
parameters $c_i$ are optimized. We then took the best configuration as the next base state.
This procedure is very similar to the stochastic variational approach
\cite{varga}. The main reason for using Gaussian basis is 
that all matrix elements can be evaluated analytically.  

\begin{figure}[h h h]
\centering
\includegraphics[width=3in]{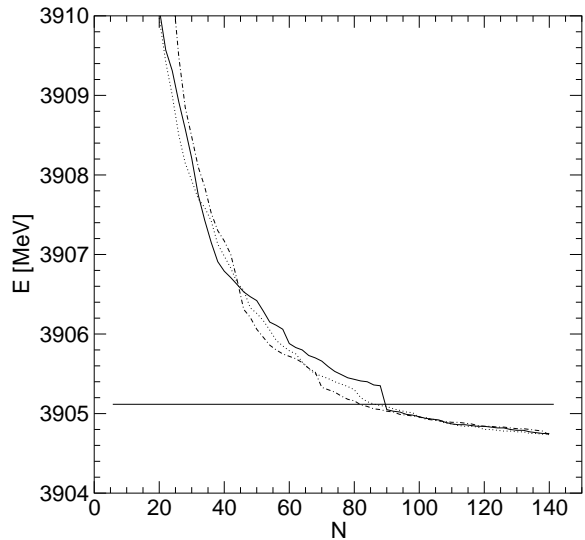}
\caption{%
Energy of the $\T_{\c\c}$ tetraquark with Bhaduri potential 
as a function of the number of the basis states for three different
runs. The $D+D^*$ threshold is also shown. Since the 
initial parameters are chosen randomly, the convergence is similar as
with the stochastic variational approach.
}
\label{slkonvergence}
\end{figure}

We were very careful that the basis states are linearly
independent so that the eigenvalues of the overlap matrix 
$\langle \psi_\alpha|\psi_\beta\rangle$ is not too close to zero which would cause numerical
instability. The dimension of the basis was between 100 (AL1 potential) and
140 (Bhaduri potential). Convergence of the energy of the $\T_{\c\c}$
tetraquark for three different runs of the code is shown in Fig.\ref{slkonvergence}. 
Here the asymptotic state of two free meson presents local minima
toward which the results are converging at first. Only at sufficiently large number
of basis states ($N>70$) the bound state can be recognized.
The initial parameters are always randomly chosen and then the optimization 
by Newton or simplex method is performed.

\end{document}